\documentclass[journal]{IEEEtranTIE}
\usepackage{graphicx}
\usepackage{cite}
\usepackage{picinpar}
\usepackage{amsmath}
\usepackage{url}
\usepackage{balance}
\usepackage{colortbl}
\usepackage{soul}
\usepackage{multirow}
\usepackage{pifont}
\usepackage{color}
\usepackage{alltt}
\usepackage[hidelinks]{hyperref}
\usepackage{enumerate}
\usepackage{siunitx}
\usepackage{epstopdf}
\usepackage{pbox}
\usepackage{stfloats}
\usepackage{booktabs}

\usepackage{graphicx}
\usepackage{subfigure}
\usepackage{verbatim}

\begin{document}

\title{Reconstructing normal section profiles of 3D revolving structures via pose-unconstrained multi-line structured-light vision}

%
%
%
%

\author{
	\vskip 1em
	{
	 Junhua Sun$^1$, Zhou Zhang$^1$, Jie Zhang$^{2*}$

	$^1$School of Instrumentation and Optoelectronic Engineering, Beihang University, Beijing 100191, China.
	$^2$School of Computer and Information Engineering, Beijing Technology and Business University, Beijing 102488, China.
	}

		
		
		
}

\maketitle

\begin{abstract}
The wheel of the train is a 3D revolving geometrical structure. Reconstructing the normal section profile is an effective approach to determine the critical geometric parameter and wear of the wheel in the community of railway safety. The existing reconstruction methods typically require a sensor working in a constrained position and pose, suffering poor flexibility and limited view-angle. This paper proposes a pose-unconstrained normal section profile reconstruction framework for 3D revolving structures via multiple 3D general section profiles acquired by a multi-line structured light vision sensor. First, we establish a model to estimate the axis of 3D revolving geometrical structure and the normal section profile using corresponding points. Then, we embed the model into an iterative algorithm to optimize the corresponding points and finally reconstruct the accurate normal section profile. We conducted real experiment on reconstructing the normal section profile of a 3D wheel. The results demonstrate that our algorithm reaches the mean precision of $0.068 mm$ and good repeatability with the STD of $0.007mm$. It is also robust to varying pose variations of the sensor. Our proposed framework and models are generalized to any 3D wheel-type revolving components.
\end{abstract}

\begin{IEEEkeywords}
normal section profile, 3D revolving structure, wheel of train, profile modeling, multi-line structured light.
\end{IEEEkeywords}


\definecolor{limegreen}{rgb}{0.2, 0.8, 0.2}
\definecolor{forestgreen}{rgb}{0.13, 0.55, 0.13}
\definecolor{greenhtml}{rgb}{0.0, 0.5, 0.0}

\section{Introduction}

\IEEEPARstart{W}{heel} of the train is a long-term running component in the railway traffic system. Periodic operation and friction gradually wear down the shape of the wheel and thus increase the potential risk of railway operation. Efficient and accurate 3D shape monitoring for the wheel is an important task in the community of railway transportation, especially with the popularization of high-speed railway vehicles.

Geometrically, the wheel is a 3D revolving body. The shape can be represented by an essential geometrical element -- normal section profile. Reconstructing the normal section profile is an intermediate approach to estimate the 3D shape of a 3D revolving structure. According to the profile reconstruction principle, the existing profile reconstruction approaches can be categorized as contact and non-contact methods. Contact profile reconstruction methods, such as MiniProf\cite{gronskov1994apparatus}, are usually time-consuming and need qualified users to work with the instrument. In the non-contact methods, Structured Light Vision (SLV) methods are representative and demonstrate high accuracy and efficiency\cite{zhang2018high,van2016real,abu2016simple}. They have been widely used for reconstructing 3D surface \cite{dong2019situ} and measuring the normal section profile\cite{molleda2016profile,wang2018distortion} as well as size parameters of the wheel\cite{torabi2018high,pan2019site}. However, it is a challenging task for the SLV sensor to reconstruct the normal section profile in a pose-unconstrained condition. We explain the problem below.

\begin{figure}[htbp]\centering
	\includegraphics[width=9cm]{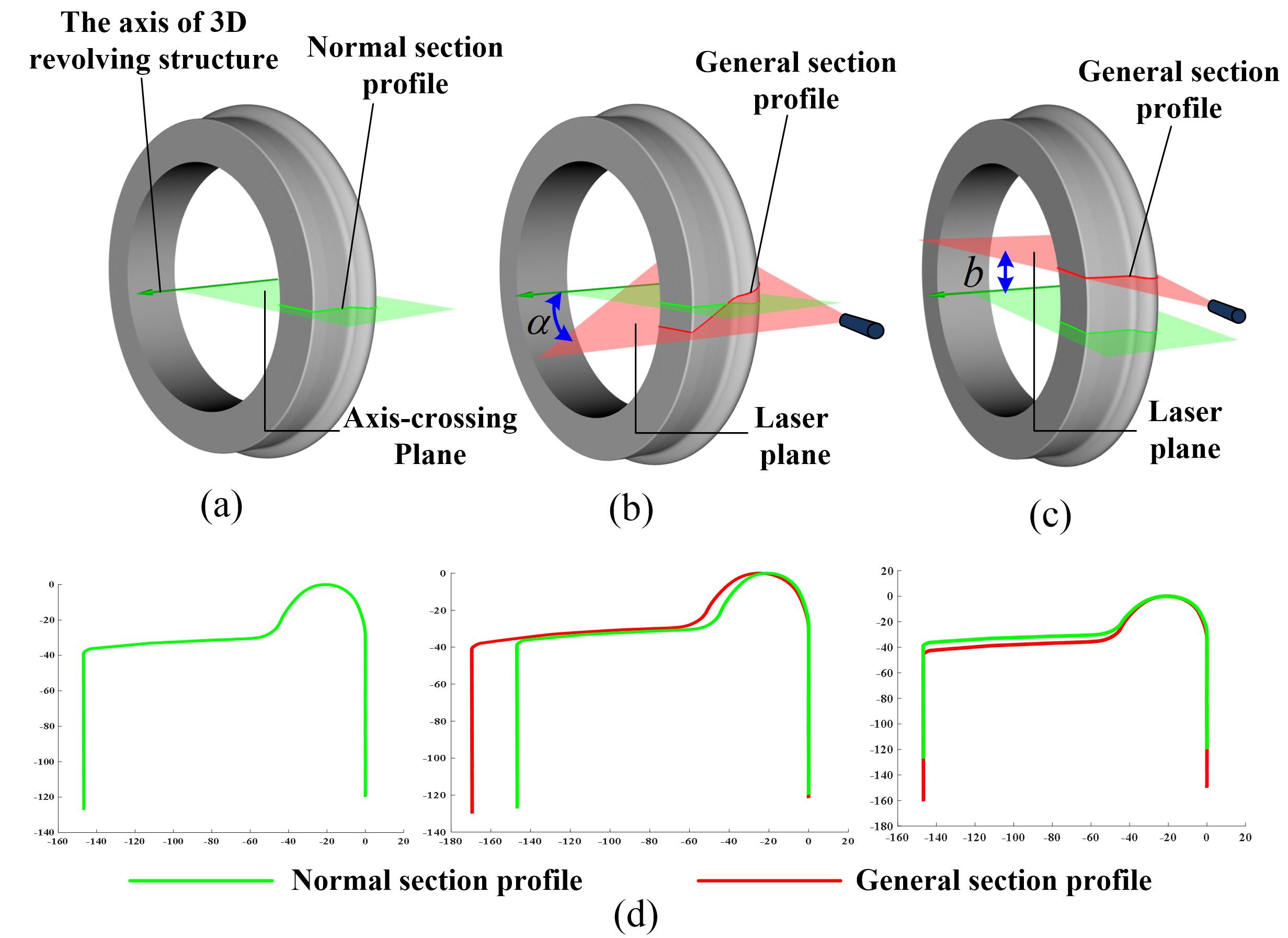}
	\caption{Challenges of normal section profile reconstruction: (a) the normal section profile of a 3D wheel-type revolving geometrical structure is the intersection of the axis-crossing plane and the 3D surface; (b) a section profile obtained by a laser plane with an off-angle $\alpha$ from the normal section profile; (c) another captured section profile with an offset $b$; (d) corresponding section profiles from the above 3 poses (the green curve is the expected normal section profile, while the actual profiles obtained by the sensor are the red ones).}
\label{Fig. 1}
\end{figure}

The normal section profile, as illustrated in \mbox{Fig. \ref{Fig. 1}(a)}, refers to the intersection of the axis-crossing plane and the 3D surface. However, a SLV sensor can only obtain the 3D section profiles at the intersection of laser plane and the 3D surface. When the laser plane is not aligned with the axis-crossing plane, the captured 3D section profile drifts from the normal one with either an off-angle or a bias. We call it a general section profile, as shown in \mbox{Fig. \ref{Fig. 1}(b)} -\mbox{Fig. \ref{Fig. 1}(c)}. Therefore, for reconstructing the normal section profile, a SLV sensor typically collaborates with an auxiliary device that fixes the sensor in a specific pose and position.

To address this issue, we explore the geometrical principle and mathematical modeling of the 3D revolving structure, and thus propose a normal section profile reconstruction algorithm for 3D revolving geometrical structures via a pose-unconstrained multi-line structured light sensor. The algorithm reconstructs the normal section profile using multiple general section profiles and allows the overall procedure to be conducted in a convenient and flexible way. Specifically, our contributions are as follows.

\begin{enumerate}
\item We establish a model to estimate the axis of the 3D rotating structure and the normal section profile using the corresponding points from multiple general section profiles. A corresponding closed-form solution for this model is derived as well.

\item Based on the above model, we propose a comprehensive algorithm
 that formulates the pose-unconstrained profile reconstruction problem into an optimization framework, where the accurate corresponding points and the normal section profile can be obtained in an iterative mode.
\end{enumerate}

As a result, the accurate normal section profile can be determined when the correspondence error converges to a minimum. Our reconstruction algorithm is fully free from pose and position constraints of the sensor.

\section{Related work}

\begin{figure*}[bp]\centering
	\includegraphics[width=17cm]{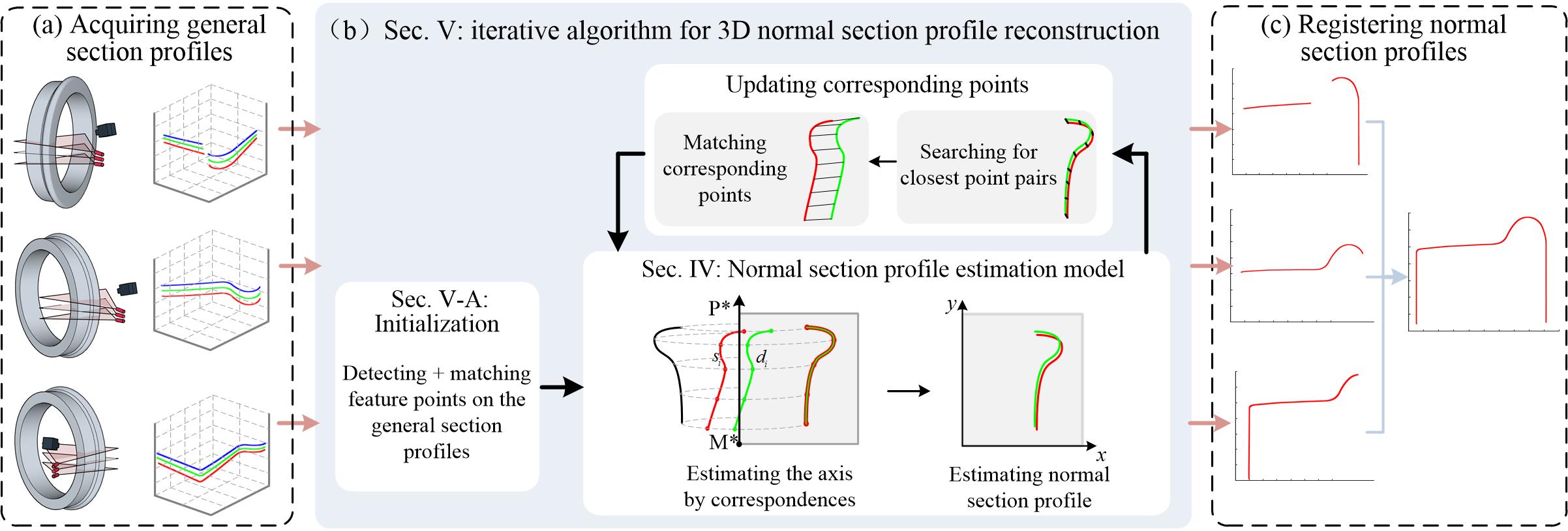}
	\caption{Overview of the proposed framework. (a) Acquiring general section profiles at different viewpoints; (b) iterative optimization algorithm for each viewpoint: the proposed model and algorithm work iteratively for matching corresponding points, estimating the axis with corresponding points and calculating normal section profile with the axis; (c) registering multiple partial normal section profiles to get a complete normal section profile.}\label{Fig. 2}
\end{figure*}

Motivated by the challenging task of reconstructing the normal section profile of the wheel, we generalize the task to a geometrical issue -- reconstructing the normal section profile of a 3D revolving geometrical structure. This section reviews the relevant work from the motivation point and focuses more on the normal section profile reconstruction approaches via the structured light vision sensor.

We categorize the existing methods as alignment-based ones and estimation-based ones. The alignment-based methods typically require the light plane to be aligned with the axis. Ref. \cite{zhang2017detection} focuses on acquiring the normal section profile of a wheel. It firstly fixed a sensor at a specific view-angle, and then the surface of the wheel was scanned continuously while rotating the wheel. The section profile with the smallest surface height (from the center) was finally selected as the normal section profile. The accuracy of this method is greatly affected by the installation accuracy of the sensor. There are also some approaches \cite{nayebi2010method,mian2009wheel,medianu2014system} using the hand-held sensor with an auxiliary alignment device. However, the auxiliary device affects operational flexibility and causes incomplete
profile reconstruction due to the limited view-angle.


The estimation-based method seeks to reconstruct the normal section profile of a 3D revolving structure via multiple general section profiles. Wu et al.\cite{wu2010novel} presented a method for on-line measurement of round steel parameters with a multi-line structured light vision sensor. In their method, the spatial ellipse centers of cross section ellipses were fitted to calculate the center axis of round steel, and then the cross section ellipses were projected along the center axis to get the corresponding normal section circles. Xing et al.\cite{xing2016online} focused on a wheel. They firstly estimated the center of the 3D revolving structure via three general section profiles and then derived the normal section profile. This approach is free from constraints on the pose and position of the sensor. However, it only uses three points to determine the center of the wheel, which makes the reconstruction accuracy sensitive to noise. The off-angle of light planes also affects the accuracy. Cheng et al.\cite{cheng2016novel} combine the alignment-based and estimation-based algorithms. They fixed the laser and calibrated the off-angle of the laser plane before the estimation, while the procedure still suffers low flexibility.

Overall, normal section profile reconstruction is still a less-exploited issue. It is necessary to develop an accurate and pose-unconstrained algorithm for reconstructing the normal section profile of the 3D revolving structures.

\section{Overview}\label{Sec_2}

\mbox{Fig. \ref{Fig. 2}} overviews the proposed framework to reconstruct the complete normal section profile of a 3D revolving structure, which includes three main steps: a) general section profile acquisition, b) normal section profile reconstruction, and c) registering multiple partial normal section profiles.

Given a 3D revolving structure, a pose-unconstrained multi-line structured-light sensor is used to acquire 3D points of the general section profiles in Fig. 2(a) according to the structured-light vision model\cite{sun2009universal}. For each viewpoint, the proposed algorithm for reconstructing the normal section profile works in an iterative mode (\mbox{Section \ref{Sec_4}}).
The axis of the 3D revolving structure is estimated via the corresponding points from multiple general section profiles.
According to the estimated axis, the normal section profile can be estimated by rotating the multiple general profiles onto an axis-crossing plane. In turn, the corresponding points between general section profiles are updated according to the closest point pairs between the estimated normal section profiles. When the correspondence error converges to a minimum, the estimated normal section profiles overlap with each other and are fused to get a final normal section profile. As the profile at one view-point is partial, we register all the partial normal section profiles from multiple single viewpoints using the ICP algorithm\cite{besl1992method,chen1992object,donoso2017icp} and obtain a complete normal section profile, as shown in \mbox{Fig. \ref{Fig. 2}(c)}. We will detail our models and the iterative optimization algorithm in the following sections.

\section{Model for normal section profile estimation }\label{Sec_3}

\begin{figure}[htbp]\centering
	\includegraphics[width=8.5cm]{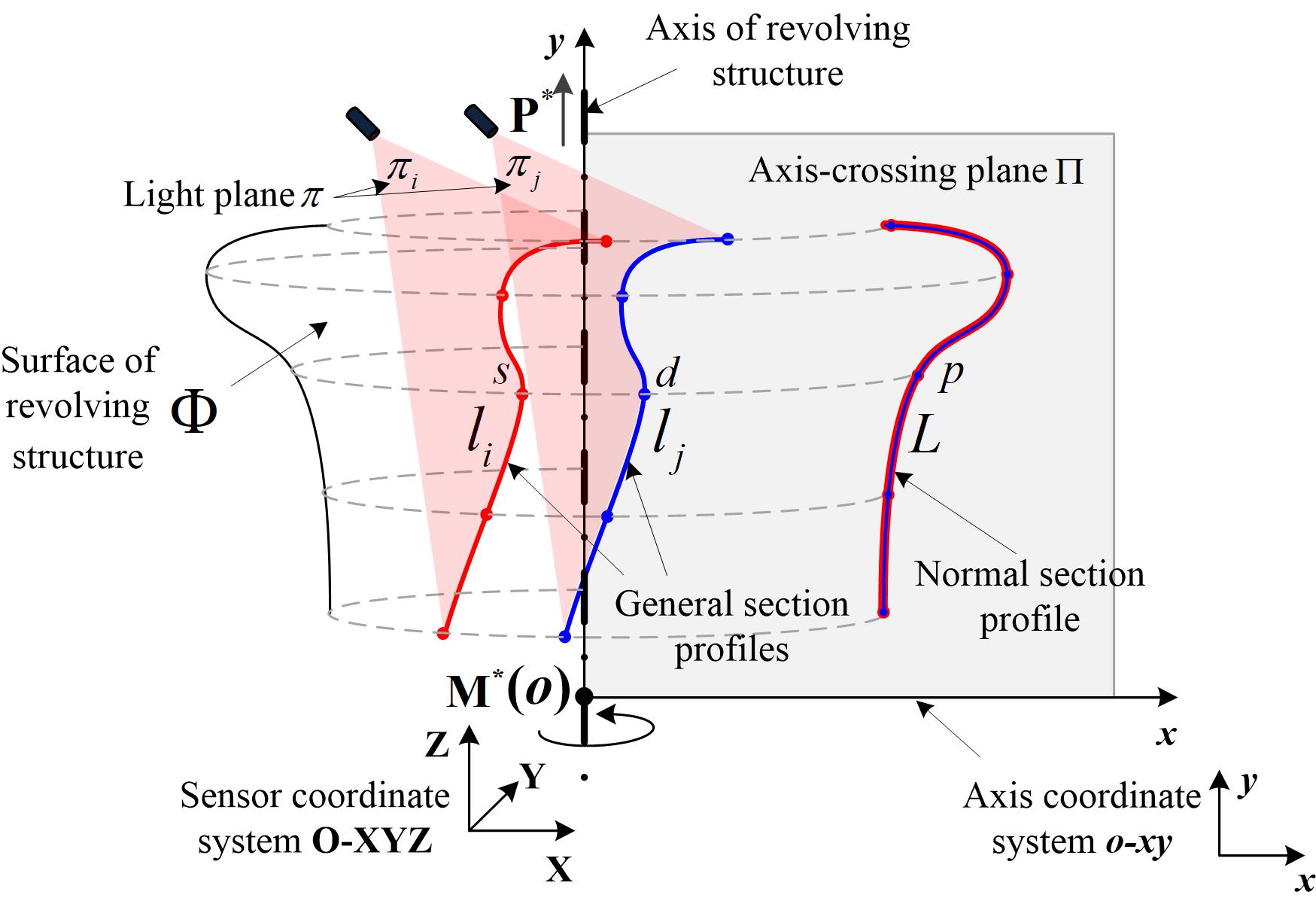}
	\caption{The model for normal section profile estimation. The position of the axis $(\mathbf{P^{*}}, \mathbf{M^{*}})$ is estimated by corresponding points $(\mathbf{s}, \mathbf{d})$ between general section profiles $l_i$ and $l_j$, and then the general section profiles are rotated around the axis and projected onto the axis-crossing plane to form the normal section profile.}\label{Fig. 4}
\end{figure}

In this section, we focus on establishing a model to estimate the axis of the 3D rotating structure and the normal section profile using the corresponding points from multiple general  section profiles. As shown in \mbox{Fig. \ref{Fig. 4}}, $\Phi$ is the surface of revolving structure,$\{\pi_k\ |\ k =1,2,...,n\}$ are the light planes and $\{l_k\ |\ k = 1,2,...,n\}$ are the general section profiles acquired by the structured light sensor ($n$ is the number of light plane). According to the structured-light vision model, $l_k$ is the set of 3D points in sensor coordinate system $\mathbf{O-XYZ}$, $l_k=\{\mathbf{q}\ |\ \mathbf{q}=(X,Y,Z), \mathbf{q}\in \Phi,\mathbf{q}\in \pi_k\}$.

The axis of revolving structure is parameterized using the axis direction and a point on the axis. In the sensor coordinate system $\mathbf{O-XYZ}$, let a unit vector $\mathbf{P}\in\mathbf{R^{3}}$ represent the axial direction and $\mathbf{M}\in\mathbf{R^{3}}$ represent a point on the axis. In order to avoid singularity, we add an additional constraint $\mathbf{P}^{T}\mathbf{M}=0$.

The normal section profile $L$ is a 2D curve on the axis-crossing plane $\Pi$ . To describe the shape of normal section profile, we establish an axis coordinate system \textbf{\emph{o-xy}} on the axis-crossing plane $\Pi$, where point $\mathbf{M}$ is the origin \textbf{\emph{o}}, the direction along $\mathbf{P}$ is \textbf{\emph{y}}-axis and the direction perpendicular to $\mathbf{P}$ is \textbf{\emph{x}}-axis. Then, the normal section profile $L$ can be described as a set of 2D points in the axis coordinate system \textbf{\emph{o-xy}}, $ L = \{  \mathbf{p}\ |\ \mathbf{p} = (x,y), \mathbf{p} \in \Phi, \mathbf{p} \in \Pi\}$.

Given the position of the axis $(\mathbf{P},\mathbf{M})$, the conversion from general section profiles $\{l_k\ |\ k = 1,2,...,n\}$ to normal section profile $ L$ can be derived. For a point $\mathbf{q}$ on general section profiles, $\mathbf{q} \in l_k$, $\mathbf{q}$ is rotated around the axis and projected onto the axis-crossing plane $\Pi$ to form a point $\mathbf{p}$ on the normal section profile $L$, $\mathbf{p} \in L$. Therefore, the conversion from $\mathbf{q}$ to $\mathbf{p}$ can be expressed as:

\begin{equation}\label{formula (1)}       
\mathbf{p}= \mathop{T}(\mathbf{q}|\mathbf{P},\mathbf{M})=
\left[            
  \begin{array}{c}   
    (\mathbf{q}-\mathbf{M})^T\mathbf{P}\\  
    |(\mathbf{q}-\mathbf{M})\times\mathbf{P}|\\  
  \end{array}
\right]                 
\end{equation}
where $(\mathbf{P},\mathbf{M})$ is the position of the axis, and $\mathop{T}(\bullet|\mathbf{P},\mathbf{M})$ is the conversion from general section profiles to the normal section profile with the given axis $(\mathbf{P},\mathbf{M})$ which is estimated by the subsequent optimization function.

In order to estimate the axis of 3D revolving structure, we first give the definition of point correspondence. As shown in \mbox{Fig. \ref{Fig. 4}}, $l_i$ and $l_j$ are any two different general section profiles in $\{l_k| k = 1,2,...,n\}$. If the axis $(\mathbf{P},\mathbf{M})$ is given exactly, $l_i$and $l_j$ overlap with the section profile $L$ after being rotated onto the axis-crossing plane by \mbox{Eq. (\ref{formula (1)})}. Therefore, for a point $\mathbf{s}\in\mathbf{R^3}$ on the general section profile $l_i$ , there is a corresponding point $\mathbf{d}\in\mathbf{R^3}$ on $l_j$. We call such a point pair $(\mathbf{s}, \mathbf{d})$ as a point correspondence, which satisfies

\begin{equation}\label{formula (0)}       
	\mathop{T}(\mathbf{s}|\mathbf{P},\mathbf{M}) = \mathop{T}(\mathbf{d}|\mathbf{P},\mathbf{M})
	\end{equation}

We assume $S$ as the set of corresponding points between general section profiles:



\begin{equation}\label{formula (2)}
S=\left\{(\mathbf{s},\mathbf{d})|\mathbf{s}\in l_{i},\mathbf{d}\in l_{j},(\mathbf{s},\mathbf{d})\ \text{satisfies}\ \mbox{Eq. (\ref{formula (0)})} \right\}
\end{equation}
where $l_{i}$ and $l_{j}$ are any two different general section profiles in $\{l_k\ |\ k = 1,2,...,n\}$.

Based on the set $S$, the axis of 3D revolving structure can be estimated by the optimization function:

\begin{equation}\label{formula (3)}
(\mathbf{P^{*}},\mathbf{M^{*}})=
\mathop{\arg\min}_{\mathbf{P},\mathbf{M}}
\sum_{(\mathbf{s},\mathbf{d})\in S}
\|T(\mathbf{s}|\mathbf{P},\mathbf{M})-T(\mathbf{d}|\mathbf{P},\mathbf{M})\|^{2}
\end{equation}

Combining \mbox{Eq. (\ref{formula (1)})}, \mbox{Eq. (\ref{formula (3)})} can be rewritten as \mbox{Eq. (\ref{formula (4)})}:

\begin{equation}\label{formula (4)}
\begin{split}
(\mathbf{P^{*}},\mathbf{M^{*}})=
\mathop{\arg\min}_{\mathbf{P},\mathbf{M}}
\sum_{(\mathbf{s},\mathbf{d})\in S}
\|\mathbf{P}^{T}(\mathbf{s} - \mathbf{d}) \|^{2}+\\
\sum_{(\mathbf{s},\mathbf{d})\in S}
\big\|\|(\mathbf{s}-\mathbf{M})\times\mathbf{P}\| - \|(\mathbf{d} - \mathbf{M})\times\mathbf{P}\| \big\|^{2}
\end{split}
\end{equation}

However, \mbox{Eq. (\ref{formula (4)})} is a nonlinear least squares problem about $(\mathbf{P}, \mathbf{M})$, which is time-consuming to be solved directly. It is necessary to derive a fast solution. The following approximation is performed to convert the problem into a combination of two linear least squares problems.

\begin{equation}\label{formula (5)}
\mathbf{P^{*}}=
\mathop{\arg\min}_{\mathbf{P}}
\sum_{(\mathbf{s},\mathbf{d})\in S}
\|\mathbf{P}^{T}(\mathbf{s} - \mathbf{d}) \|^{2}
\end{equation}

\begin{equation}\label{formula (6)}
\mathbf{M^{*}}=
\mathop{\arg\min}_{\mathbf{M}}
\sum_{(\mathbf{s},\mathbf{d})\in S}
\big\|\|(\mathbf{s}-\mathbf{M})\times\mathbf{P^{*}}\| - \|(\mathbf{d} - \mathbf{M})\times\mathbf{P^{*}}\| \big\|^{2}
\end{equation}

For the first linear least squares problem in \mbox{Eq. (\ref{formula (5)})}, we add the constraint $\|\mathbf{P}\|=1$ and convert it into the standard form of linear least squares as \mbox{Eq. (\ref{formula (7)})}.

\begin{equation}\label{formula (7)}
\begin{split}
&\mathbf{D}\mathbf{P}=0\textrm{ , }
\mathbf{D}=
\left[
  \begin{array}{c}   
    (\mathbf{s}_{1}-\mathbf{d}_{1})^T\\  
    \vdots \\  
    (\mathbf{s}_{N}-\mathbf{d}_{N})^T\\  
  \end{array}
\right]\\
&s.t.\|\mathbf{P}\|=1
\end{split}
\end{equation}\\
where $N$ is the number of correspondences in $S$.

Thereby, $\mathbf{P}^{*}$ is the feature vector corresponding to the minimum eigenvalue of the matrix $\mathbf{D}^T\mathbf{D}$.

For the second least squares problem in \mbox{Eq. (\ref{formula (6)})}, it is also a linear least squares problem about point $\mathbf{M}$ since the optimal axial direction $\mathbf{P}^{*}$ have been estimated by \mbox{Eq. (\ref{formula (5)})}. Here, we let $[\mathbf{P^{*}}]_{\times}$ represent the anti-symmetric matrix of $\mathbf{P}^{*}=(p_{1},p_{2},p_{3})^{T}$,
$[\mathbf{P^{*}}]_{\times} = \begin{bmatrix}
0 & -p_{3} & p_{2}\\
p_{3} & 0 & -p_{1}\\
-p_{2} & p_{1} & 0\\
\end{bmatrix}$. Then the linear equations about $\mathbf{M}$ can be given as \mbox{Eq. (\ref{formula (8)})}.

\begin{equation}\label{formula (8)}
\mathbf{H}_{i}^{T}\mathbf{M}=b_{i}\textrm{ , }
i=1,2\dots N
\end{equation}\\
where

\begin{equation}
\mathbf{H}_{i} = [\mathbf{P^{*}}]^{T}_{\times}[\mathbf{P^{*}}]_{\times}(\mathbf{s}_i-\mathbf{d}_i)
\end{equation}\\

\begin{equation}
b_{i}=\frac{1}{2}(\mathbf{s}_{i}+\mathbf{d}_{i})^{T}[\mathbf{P^{*}}]^{T}_{\times}[\mathbf{P^{*}}]_{\times}(\mathbf{s}_i-\mathbf{d}_i)
\end{equation}\\

Adding the constraint $\mathbf{P}^{*T}\mathbf{M}=0$, \mbox{Eq. (\ref{formula (8)})} is converted into the standard form of linear least squares about $\mathbf{M}$.

\begin{equation}\label{formula (9)}
\begin{split}
&\mathbf{H}\mathbf{M}=\mathbf{B}\textrm{ , }
\mathbf{H}=
\left[
  \begin{array}{c}   
    \mathbf{H}_{1}^{T}\\  
    \vdots \\  
    \mathbf{H}_{N}^{T}\\  
    \mathbf{P}^{*T}
  \end{array}
\right]\textrm{ , }
\mathbf{B}=
\left[
  \begin{array}{c}   
    b_{1}\\  
    \vdots \\  
    b_{N}\\  
    0\\
  \end{array}
\right]
\end{split}
\end{equation}

Finally, $\mathbf{M}^{*}$ can be calculated by:

\begin{equation}\label{formula (10)}
\mathbf{M}^{*} = (\mathbf{H}^{T}\mathbf{H})^{-1}\mathbf{H}^{T}\mathbf{B}
\end{equation}

After the optimal estimate $(\mathbf{P^*},\mathbf{M^*})$ of the axis is calculated via the optimization function in \mbox{Eq. (\ref{formula (3)})}, the general section profiles can be rotated around the axis and projected onto the axis-crossing plane to form the normal section profiles:

\begin{equation}\label{formula (normal)}
L_k = \{\mathbf{p}\ |\ \mathbf{p}=T(\mathbf{q}|\mathbf{P^*},\mathbf{M^*}), \mathbf{q} \in l_k\}
\end{equation}
where $L_k$ is the normal section profile converted from general section profile $l_k \in \{l_k\ |\ k=1,2...n\}$, $T(\bullet|\mathbf{P^*},\mathbf{M^*})$ is the conversion from a general section profile to the normal section profile in \mbox{Eq. (\ref{formula (1)})}.

Overall, a normal section profile estimation model based on point correspondences is derived in this section. \mbox{Eq. (\ref{formula (3)})} is the optimization function for estimating the position of axis via point correspondences $S$. \mbox{Eq. (\ref{formula (7)})} and \mbox{Eq. (\ref{formula (10)})} give closed-form solutions to $\mathbf{P}$ and $\mathbf{M}$ respectively. \mbox{Eq. (\ref{formula (1)})} is the conversion from general section profiles to the normal section profile via the position of axis $(\mathbf{P},\mathbf{M})$.

\section{Iterative algorithm for reconstructing normal section profile}\label{Sec_4}

In order to reconstruct the normal section profile from general section profiles via the model in \mbox{Section \ref{Sec_3}}, the exact correspondences $S$ must be obtained. This section goes through the proposed comprehensive reconstruction framework, which starts from the initial correspondences and iterates between matching more accurate correspondences and calculating more accurate normal section profile. The proposed framework consists of two parts: feature point detection and initial matching (\ref{featuredete}) and iterative optimization (\ref{optimization}).

\subsection{Feature point detection and initial matching}
\label{featuredete}
\begin{figure}[htbp]\centering
	\includegraphics[width=9cm]{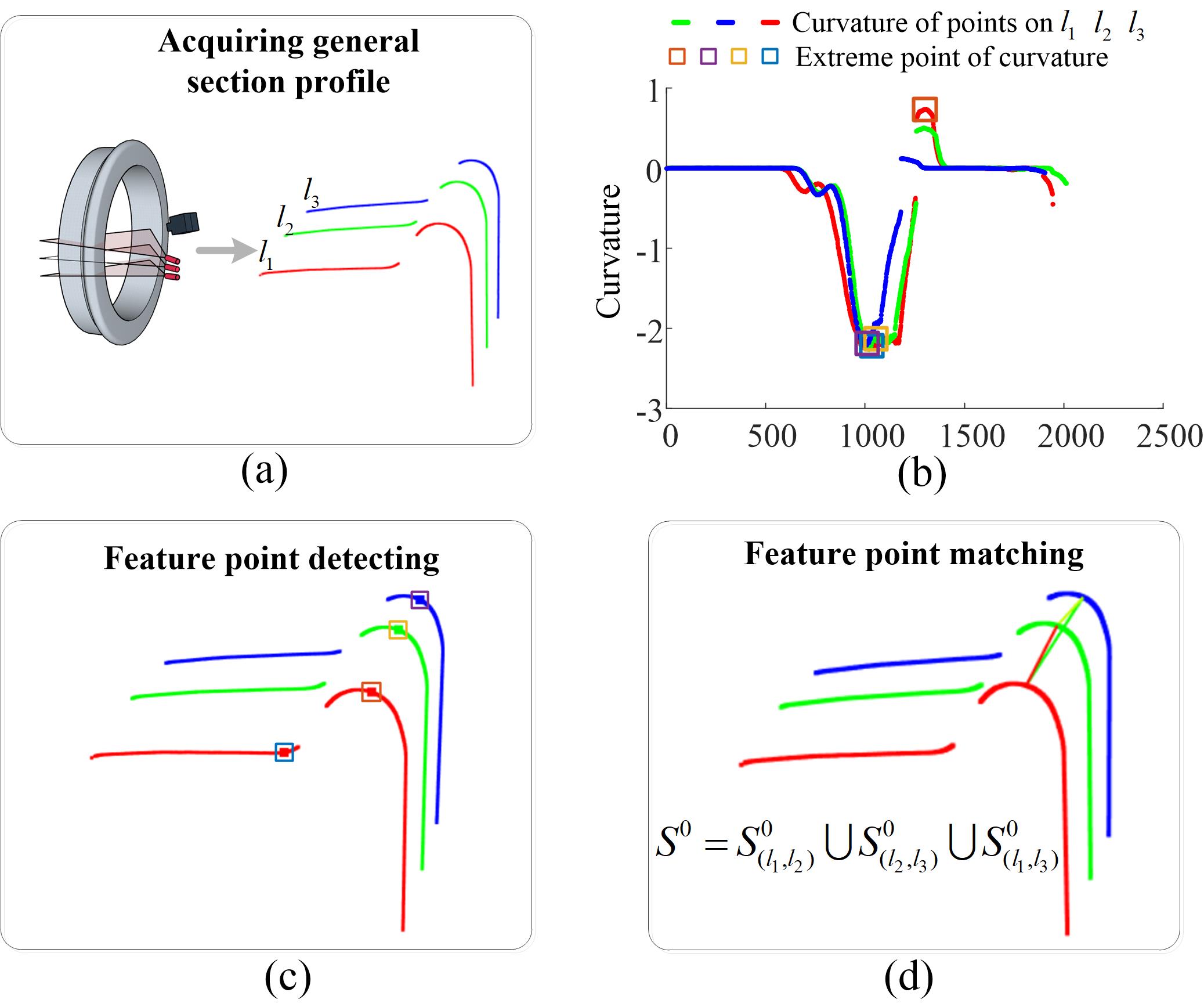}
	\caption{Feature point detecting and matching: (a) acquiring general section profiles $\{l_1,l_2,l_3\}$; (b) calculating curvature value point by point and extracting the local extreme of curvature sequence; (c) detecting the points with extremum of curvature as feature points; (d) matching feature points to get initial corresponding points set $S^{0}$. (The number of general section profiles $n = 3$ is taken as an example)}\label{Fig. 5a}
\end{figure}

Feature point detection and matching provide initial set of corresponding points $S^{0}$, which is showed in \mbox{Fig. \ref{Fig. 5a}}. We extract the extreme points of curvature\cite{pottmann2009integral} as feature points and match these feature points as corresponding points by \mbox{Eq. (\ref{formula (11)})}.

\begin{equation}\label{formula (11)}
\mathop{dis}(p,q) = \frac{|C_{p}-C_{q}|}{\max\{C_{p},C_{q}\}}
\end{equation}\\
where $p$ and $q$ are feature points on different general section profiles, $C_p$ and $C_q$ are curvature values of $p$ and $q$ respectively.

The value of $\mathop{dis}(p,q)$ in \mbox{Eq. (\ref{formula (11)})} describes the matching distance between two feature points. We first set a threshold for the matching distance to select correspondence candidates and then use a mean shift clustering algorithm\cite{cheng1995mean} to determine the final correspondences. By matching the feature points between each two general section profiles $(l_1,l_2)$, $(l_2,l_3)$ and $(l_3,l_1)$, the initial corresponding point set $S^{0}=S^{0}_{(l_1,l_2)}\cup S^{0}_{(l_2,l_3)}\cup S^{0}_{(l_3,l_1)}$ can be obtained.

\subsection{Iterative optimization between correspondences and normal section profile} \label{Sec_It}
\label{optimization}
\begin{figure*}[htbp]\centering
	\includegraphics[width=18cm]{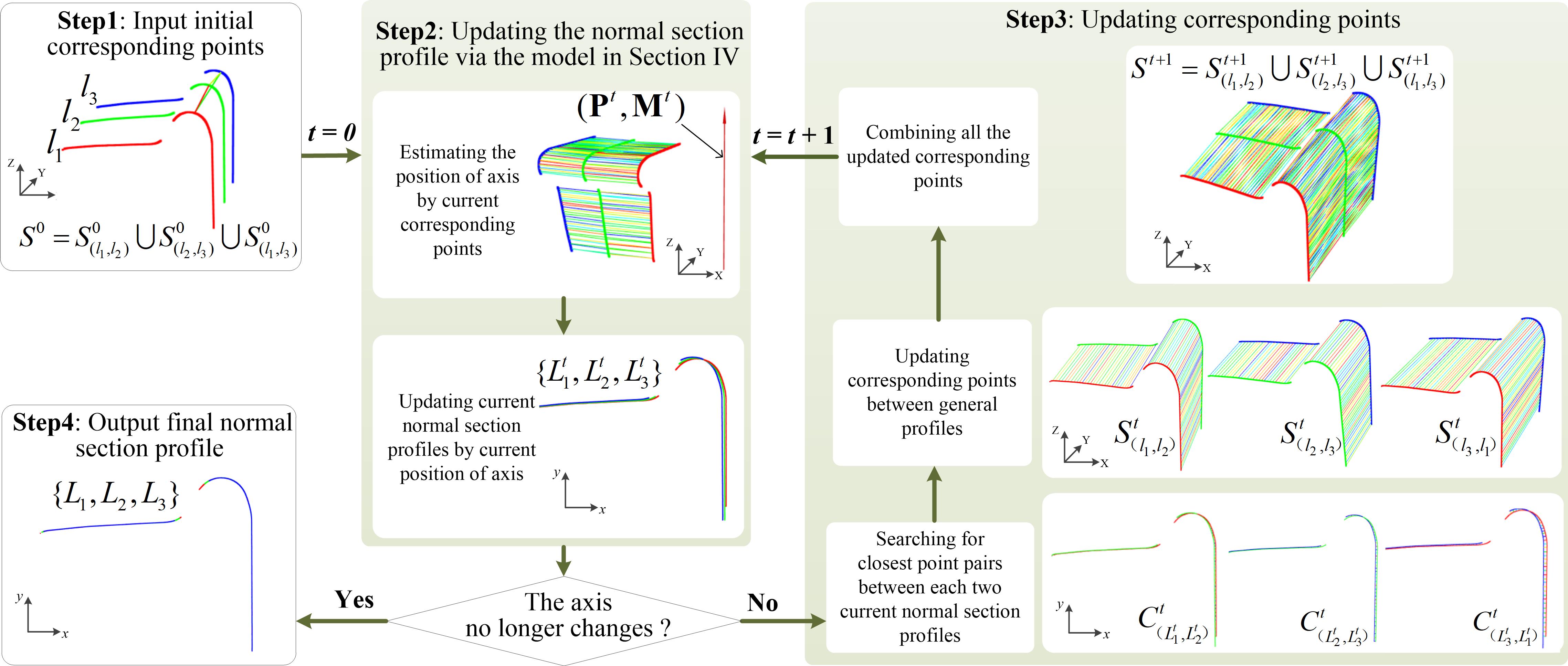}
	\caption{The iterative optimization process (Please refer to \mbox{Section \ref{Sec_It}} for the explanation of every step). $\mathbf{O-XYZ}$ is the sensor coordinate system and $\mathbf{o-xy}$ is the axis coordinate system on the axis-crossing plane.}\label{Fig. 5b}
\end{figure*}

As shown in \mbox{Fig. \ref{Fig. 5b}}, the second part of the algorithm iterates between updating normal section profile and updating correspondence set $S^t$ ($t$ is current iteration count), and outputs the final normal section profile in the end. The process in \mbox{Fig. \ref{Fig. 5b}} is detailed as follows. Here the number of general section profiles, $n = 3$ is taken as an example.

\textbf{Step1} Input the initial corresponding points set $S^{0}$ and let current iteration count $t =0$.

\textbf{Step2} Updating normal section profiles via normal section profile estimating model in \mbox{Section \ref{Sec_3}}. First, according to current correspondence set $S^{t}$, we estimate current position of axis $(\mathbf{P},\mathbf{M})^{t}$. Next, according to current position of axis $(\mathbf{P},\mathbf{M})^{t}$, the general section profiles $\{l_1,l_2,l_3\}$ are rotated around the axis and projected onto the axis-crossing plane to get current normal section profiles $\{L_{1}^{t}, L_{2}^{t},L_{3}^{t} \}$. Then, we determine whether the axis $(\mathbf{P},\mathbf{M})^{t}$ has changed compared to the previous iteration. If not, we terminate the iteration and go to Step4.

\textbf{Step3} Search for closest point pairs between each two profiles in $\{L_{1}^{t}, L_{2}^{t},L_{3}^{t} \}$ to get the set of closest point pairs $C^{t}=C^{t}_{(L_1^t,L_2^t)}\cup C^{t}_{(L_2^t,L_3^t)}\cup C^{t}_{(L_3^t,L_1^t)}$. According to the indexes of closest points on current normal section profiles $\{L_{1}^{t}, L_{2}^{t},L_{3}^{t} \}$, we select the point pairs with same indexes on general section profiles to update correspondence set $S^{t+1}=S^{t+1}_{(l_1,l_2)}\cup S^{t+1}_{(l_2,l_3)}\cup S^{t+1}_{(l_3,l_1)}$. Let iteration count $t = t+1$, and go to Step2.

\textbf{Step4} The final normal section profiles $\{L_1,L_2,L_3\}$ from the last iteration overlap with each other and are fused to get the accurate normal section profile $L$.

\section{Experiments}\label{Sec_5}

The experiments verify the accuracy and the robustness of the proposed framework at different viewpoints when measuring a complete 3D normal section profile.

\subsubsection{Experimental setup}

\begin{figure}[!t]\centering
	\includegraphics[width=9cm]{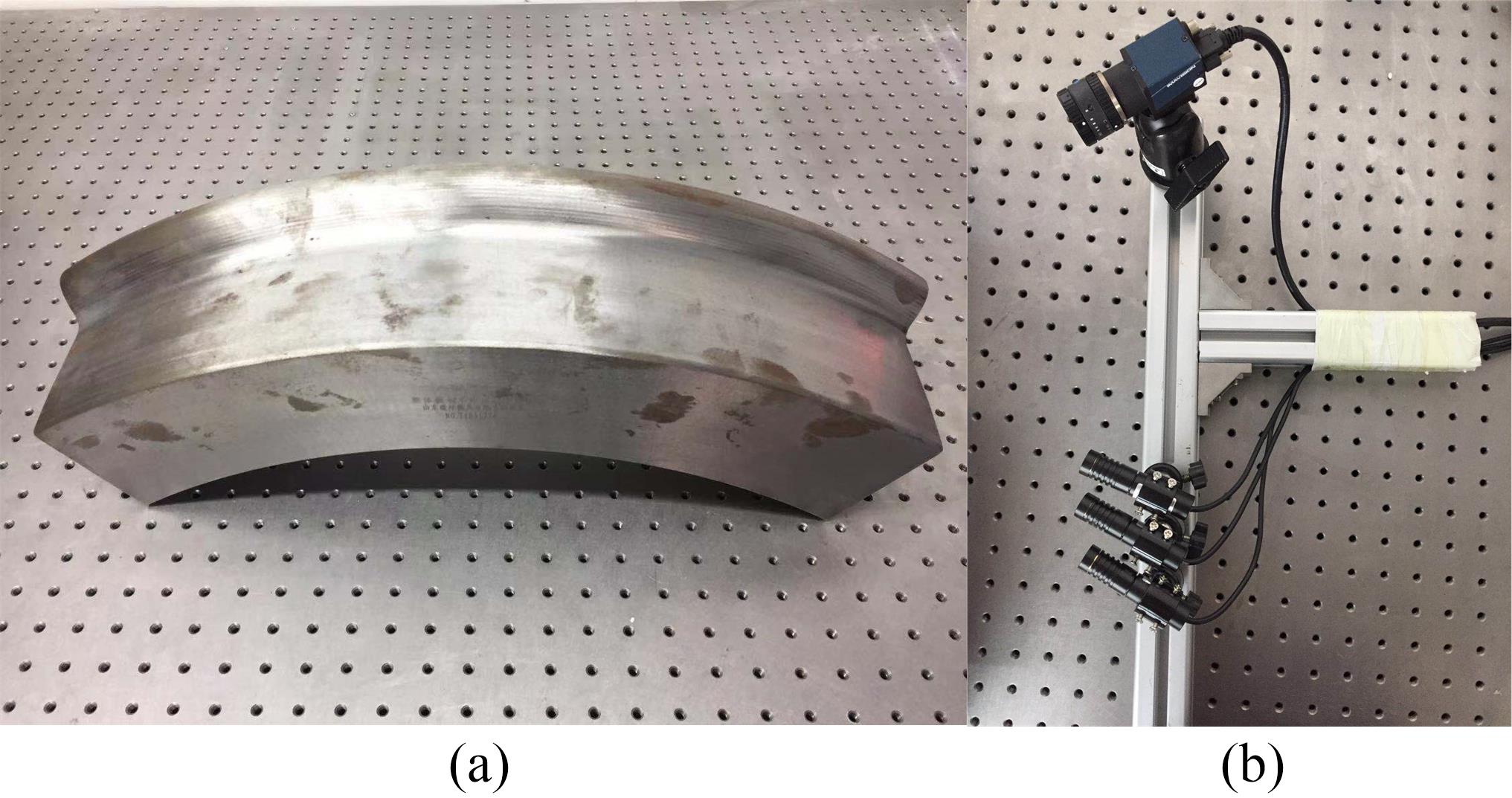}
	\caption{Experiment equipment: (a) a partial wheel of the train; (b) a pose-unconstrained multi-line structured light system, with 3 lasers and a camera installed on a hand-held trestle.}\label{Fig. 10}
\end{figure}

We used a partial wheel of the train as a 3D revolving geometrical structure and a multi-line structured light sensor for acquiring raw 3D point cloud, as shown in \mbox{Fig. \ref{Fig. 10}}. Specifically, 3 lasers and a camera were installed on a handheld trestle, and the spacing of light planes is around 20 mm. The other parameters are set as follows. The resolution of the camera is 2448 pixels $\times$ 2048 pixels, and the power of the lasers is 30 mW. The diameter of the 3D wheel is 1040 mm. The working distance (from the camera to the measured surface)of the sensor is about 400 mm.

\begin{table}[htbp]
	\centering
	\caption{The camera's internal parameters and distortion coefficients}
	\resizebox{\columnwidth}{!}{
	  \begin{tabular}{cccc}
	  \bottomrule
	  \multicolumn{2}{c}{Parameter type} & Value & Accuracy \\
	  \midrule
	  \multirow[c]{2}[0]{*}{Scale factor} & $f_x$ & 2430.062  & 0.351  \\
			& $f_y$ & 2430.067  & 0.351  \\
	  \midrule
	  \multirow[c]{2}[0]{*}{Principal point} & $u_0$/pixel & 1231.960  & 0.186  \\
			& $v_0$/pixel & 1020.465  & 0.134  \\
	  \midrule
	  Distortion  & $k_c$    & \multicolumn{2}{c}{[ -0.1062, 0.1823, 0.0001, 0.0004, 0.0000 ]} \\
	  \bottomrule
	  \end{tabular}%
	}
	\label{TABLE III}%
  \end{table}%

\begin{table}[!t]
	\centering
	\caption{The equations of light planes}
	  \begin{tabular}{cccccc}
	  \toprule
	  \multirow[c]{2}[1]{*}{light plane} & \multicolumn{4}{c}{$a_{0}x + a_{1}y + a_{2}z +a_{3} = 0$}          & \multirow[c]{2}[1]{*}{RMS/mm} \\
  \cmidrule{2-5}          &    $a_0$   &   $a_1$    &    $a_2$  &    $a_3$   &  \\
	  \midrule
	  light plane 1$_{th}$  & -0.575  & -0.027  & 0.818  & -190.756  & 0.031  \\
	  light plane 2$_{th}$  & -0.528  & -0.040  & 0.849  & -209.491  & 0.033  \\
	  light plane 3$_{th}$  & -0.481  & -0.037  & 0.876  & -227.202  & 0.031  \\
	  \bottomrule
	  \end{tabular}%
	\label{TABLE IV}%
  \end{table}%

We calibrated the camera and the three light planes using Zhang's method\cite{zhang2000flexible} and Sun's method\cite{sun2009universal} respectively. The calibration results of the camera's internal parameters and distortion coefficients are shown in \mbox{Table \ref{TABLE III}}, and the equations of light planes are shown in \mbox{Table \ref{TABLE IV}}.

\subsubsection{Reconstructing normal section profile at different viewpoints}

\begin{figure}[!t]\centering
	\includegraphics[width=9cm]{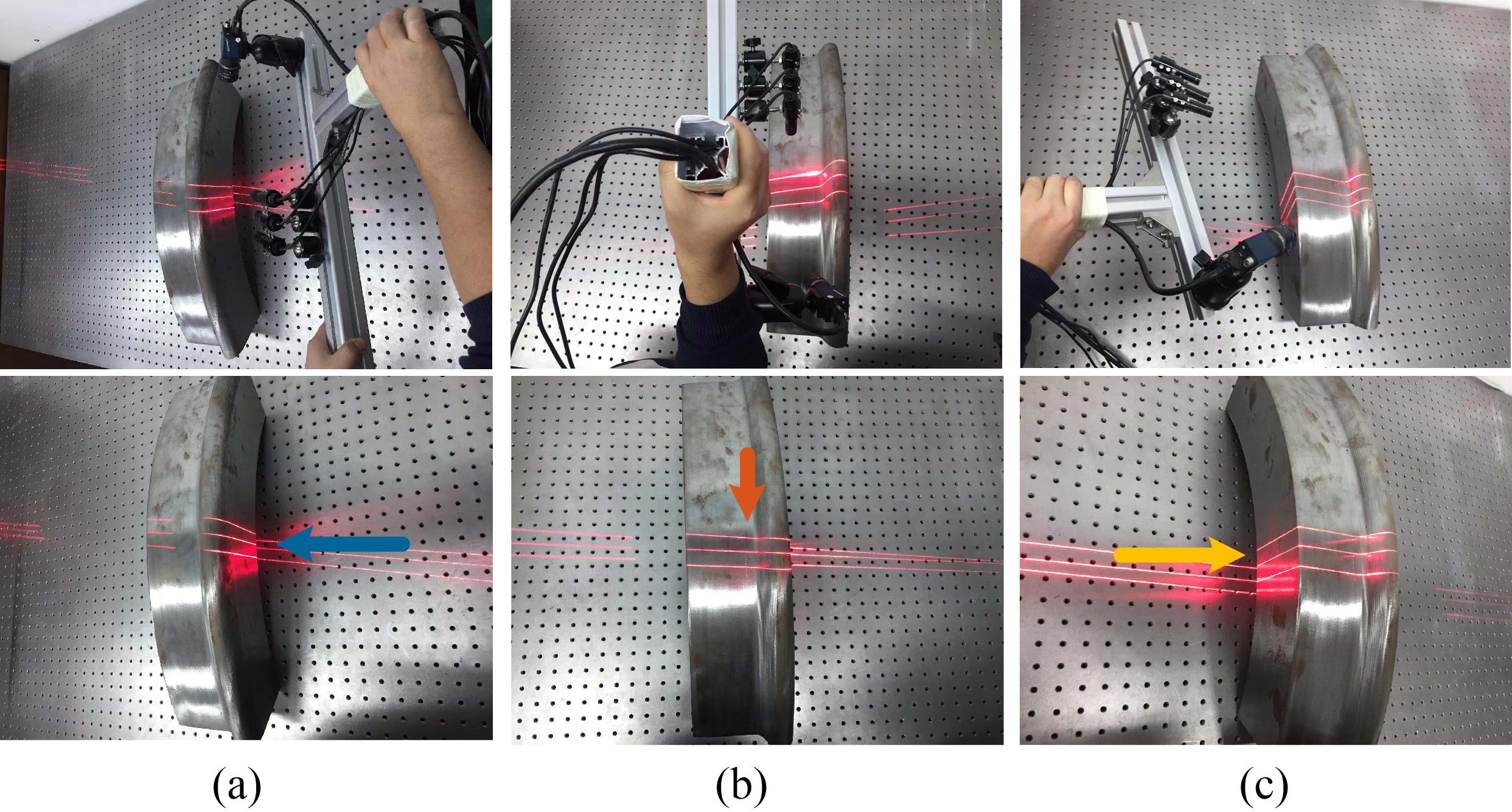}
	\caption{Three different viewpoints of the sensor: (a) light planes hit the surface from back of the flange; (b) light planes hit the surface from the wheel tread; (c) light planes hit the surface from the rim face. Different parts of the normal section profile can be reconstructed from the different viewpoints.}\label{Fig. 11}
\end{figure}

To investigate the performance of the algorithm at various viewpoints, we set the sensor to three common viewpoints, as shown in
\mbox{Fig. \ref{Fig. 11}}. We conducted the normal section profile reconstruction at each viewpoint.

We set a quantitative assessment criteria using the corresponding points in the iterative framework. That is, we calculate the Root Mean Square (RMS) of all the correspondence errors that evaluate the convergence of the algorithm, as \mbox{Eq. (\ref{formula (15)})}. When the normal section profile is reconstructed successfully, the corresponding points $({\bf{s}},{\bf{d}})$ between different general section profiles are expected to be overlapped after they are projected onto the axis-crossing plane. Correspondingly, the correspondence error is expected to be small as well.

\begin{equation}\label{formula (15)}
	e =\Big( \frac{1}{N}\sum_{(\mathbf{s},\mathbf{d})\in S}\big|\mathop{T}(\mathbf{s}|\mathbf{P},\mathbf{M})-\mathop{T}(\mathbf{d}|\mathbf{P},\mathbf{M})\big| \Big)^{\frac{1}{2}}
\end{equation}\\
where $S$ is the correspondence set, $N$ is the number of correspondences in $S$, $\mathop{T}(\bullet|\mathbf{P},\mathbf{M})$ is the conversion from general section profile to normal section profile with the axis parameters $({\bf{P},\bf{M}})$.

\begin{figure*}[htbp]\centering
	\includegraphics[width=16.5cm]{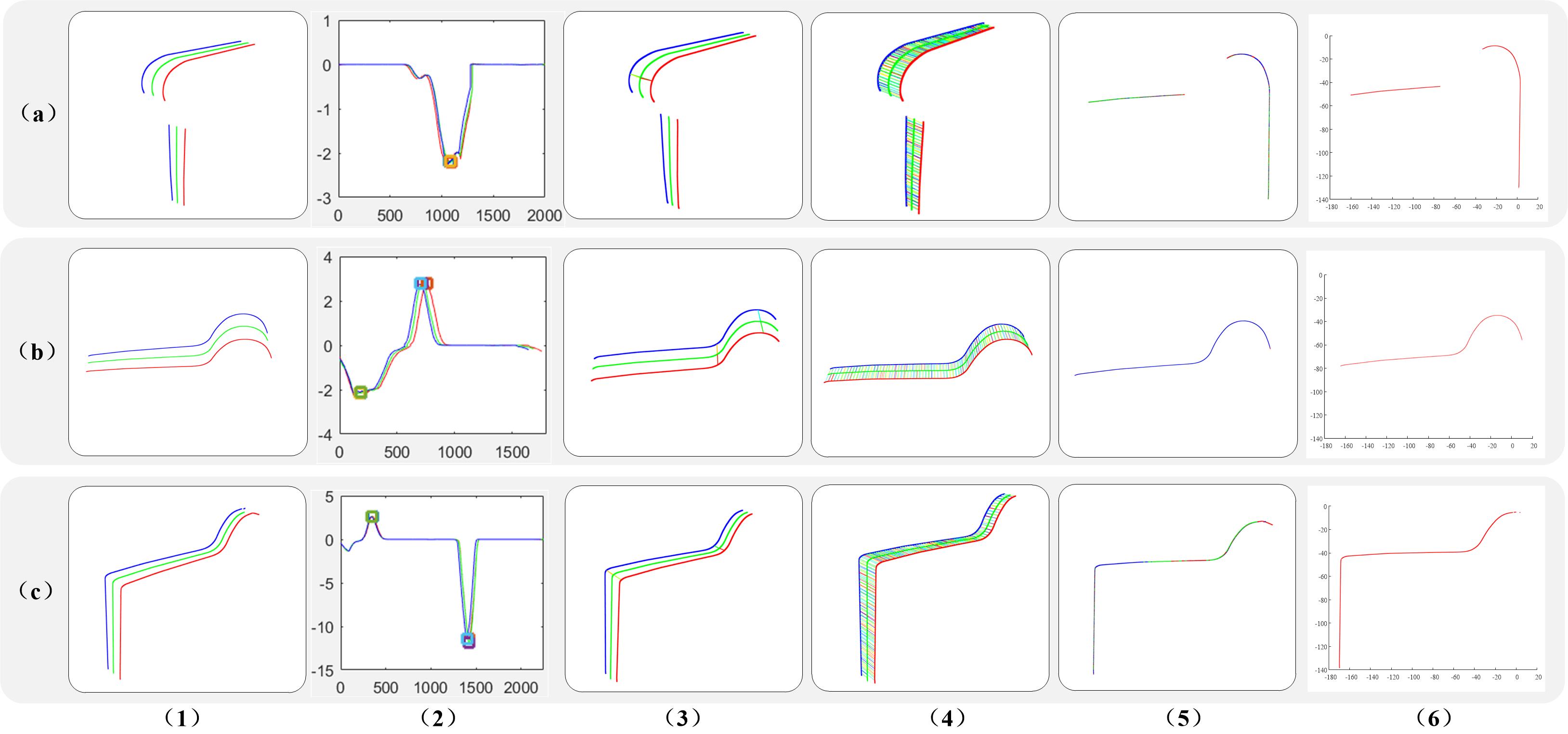}
	\caption{The intermediate results of normal section profile reconstruction. Each row shows the process of one viewpoint. (1) General section profiles obtained by the hand-held sensor; (2) curvature sequences of general section profiles, where the local extremes are extracted as feature points; (3) matching feature points on general section profiles to get the initial correspondence set $S^0$; (4) ultimate corresponding points through iterative optimization (For visualization, we only draw $5\%$ of all the correspondence); (5) the final normal section profiles after iterative optimization, which overlap on the axle-crossing plane; (6) the final reconstructed partial normal section profiles.}\label{Fig. 12}
\end{figure*}

\begin{figure*}[htbp]\centering
	\includegraphics[width=16.5cm]{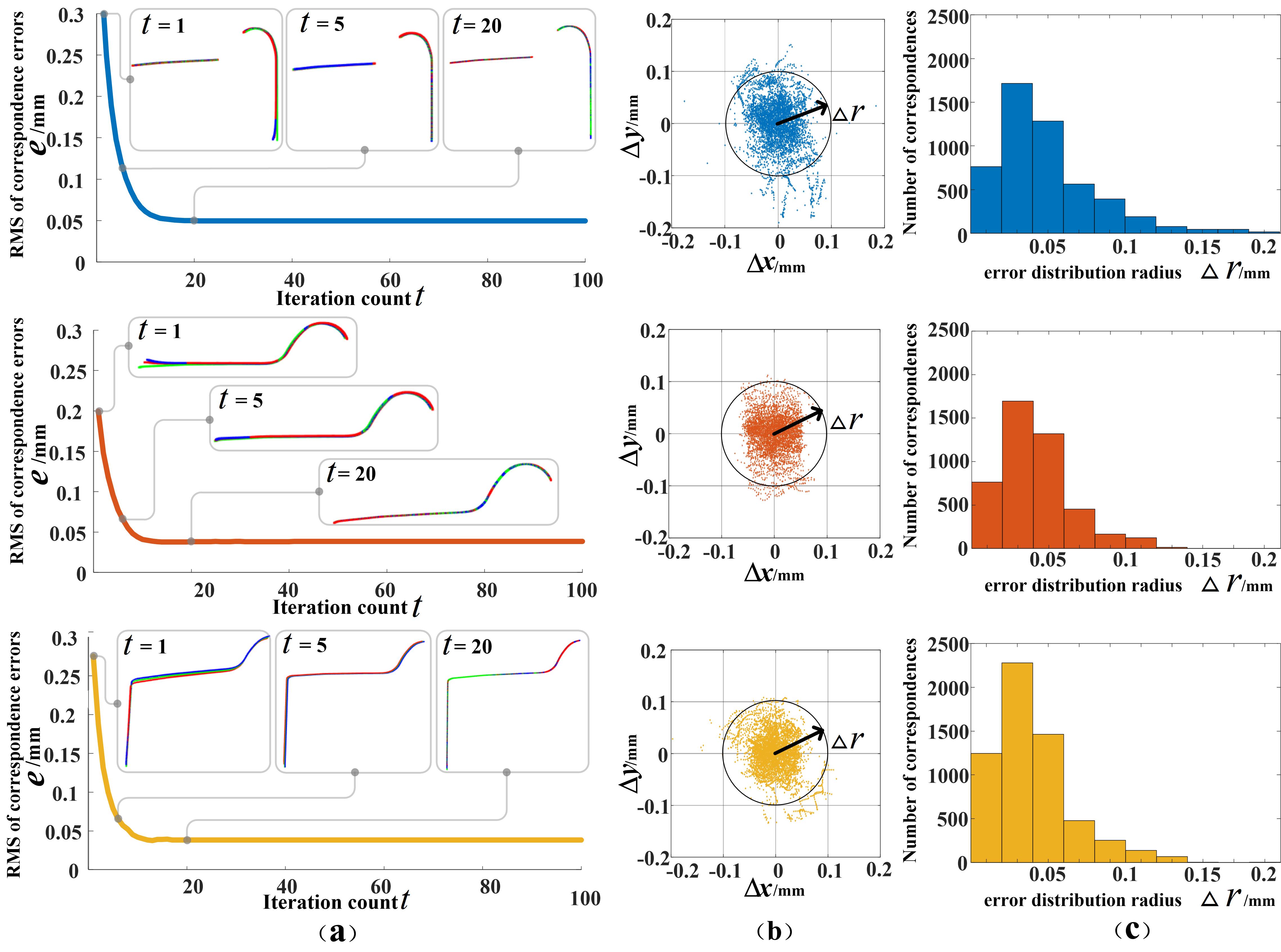}
	\caption{The convergence process and correspondence error distribution of the proposed algorithm. The three rows represent the case a-c respectively. (a) The RMS of convergence errors $e$ vs iteration count $t$. (b) The scatter plots of the correspondence errors after the algorithm converges ($t > 20$). (c) The histograms of the correspondence errors after the algorithm converges ($t > 20$).}\label{Fig. 13}
\end{figure*}

\mbox{Fig. \ref{Fig. 12}} shows the intermediate process of reconstructing normal section profile from the above 3 different viewpoints. We firstly obtained 3D general section profiles by the multi-line structured light sensor in \mbox{Fig. \ref{Fig. 12}}(1). For getting initial corresponding points from the general profiles, we extracted and matched the 3D points with curvature extremes as feature points in \mbox{Fig. \ref{Fig. 12}}(2)-(3). \mbox{Fig. \ref{Fig. 12}}(4) are the final corresponding points after iterative optimization (For clearer view, we only draw $5\%$ of all the correspondences). Based on the corresponding points, we calculated the axis of the 3D wheel and then the 3 general profiles were converted into the normal section profiles via the calculated axis. The final normal section profiles overlap with each other, as shown in \mbox{Fig. \ref{Fig. 12}}(5). Finally, we get the reconstructed partial normal section profiles from the 3 different viewpoints in \mbox{Fig. \ref{Fig. 12}}(6).

\mbox{Fig. \ref{Fig. 13}} shows the convergence process with the error distribution of correspondence errors. From Fig. 10(a), the algorithm successfully converges within 20 iterations in all the three cases. The final RMS of the correspondence errors converges to 0.05 mm roughly. It indicates that the algorithm can converge stably when dealing with different viewpoints. We also show the distribution of all the correspondence errors in Fig. 10(b). For quantitative evaluation, we use an error region with a radius $\Delta r$ to measure the number of correspondences and thus give out the statistical histograms in Fig. 10(c). We can see that most of the correspondence errors are within 0.1 mm and follow the normal distribution with the average of 0.04 mm roughly.

\subsubsection{Accuracy analysis of complete profile}

\begin{figure*}[htbp]\centering
	\includegraphics[width=15cm]{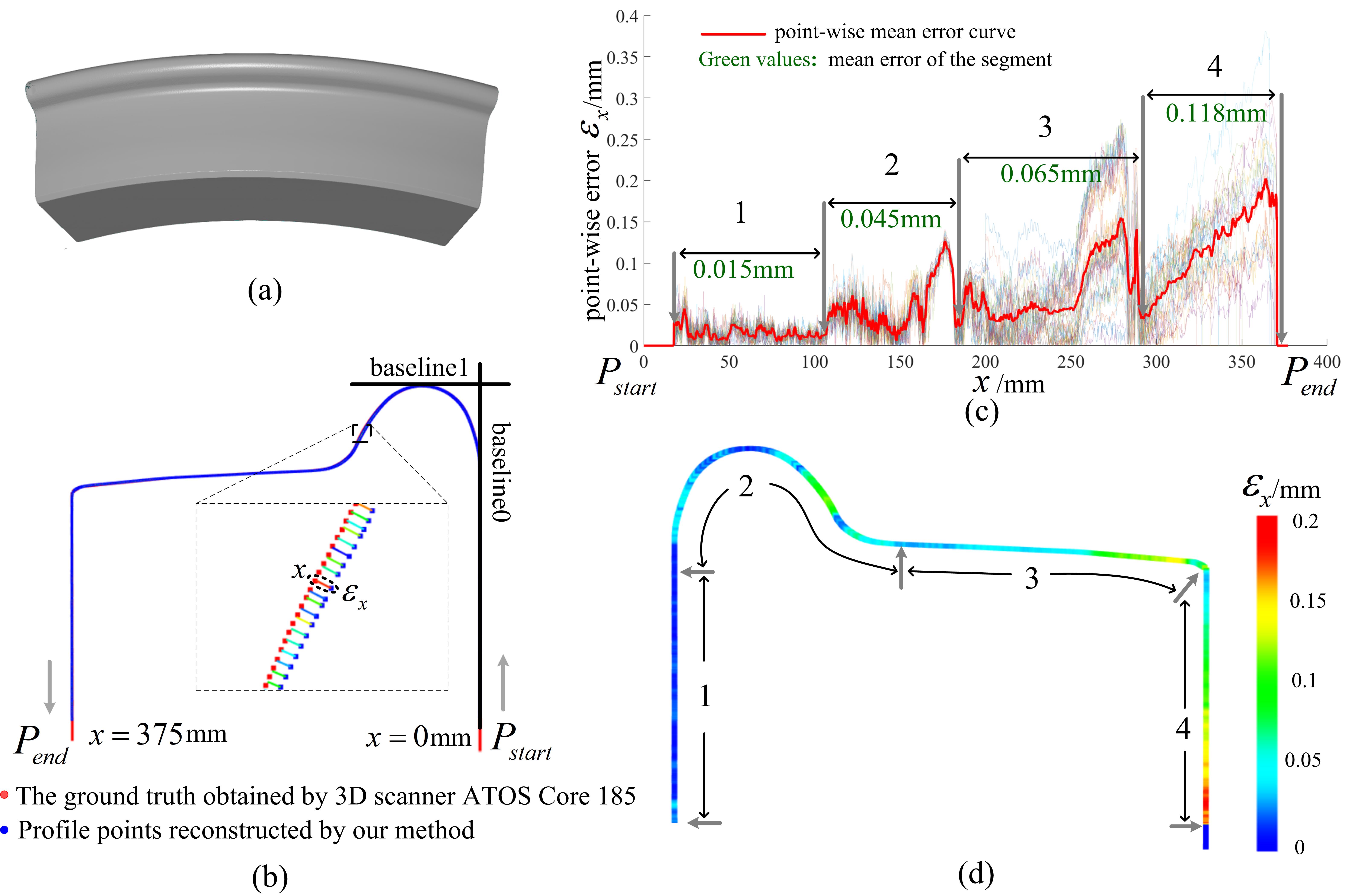}
	\caption{Accuracy analysis: (a) 3D ground truth model of the wheel obtained by a 3D scanner ATOS Core 185. (b) Comparison between the reconstructed normal section profile and the ground truth. The point-wise error $\varepsilon_x$ is calculated from $P_{start}$ to $P_{end}$, where $x$ is the arc length from a current point to $P_{start}$. (c) Repeatability: point-wise error curves in 30 repetitions. (d) Heat map of mean errors on the normal section profile.}\label{Fig. 14}
\end{figure*}

We registered the partial normal section profiles from multiple viewpoints to get a complete normal section profile and evaluate the accuracy of the complete profile. To obtain the ground truth of the normal section profile, we scanned the 3D surface of the wheel using an off-the-shelf 3D scanner -- ATOS Core 185 from GOM company \cite{ATOSCore185}, whose scanning accuracy reaches 0.014 mm according to its official document (VDI/VDE 2634 part 3). We firstly obtained the axis of the wheel by fitting the inner cylinder. Then, the axis-crossing plane intersects the wheel surface, which generates the normal section profile (as the ground truth). \mbox{Fig. \ref{Fig. 14}(a)} shows the complete scanning result of the wheel. As shown in \mbox{Fig. \ref{Fig. 14}(b)}, we aligned the reconstructed normal section profile with the ground truth according to the baseline 0 and baseline 1 and calculated the point-wise error as \mbox{Eq. (\ref{formula (16)})}.

\begin{equation}
\label{formula (16)}
	\varepsilon_x = \mathop{d}(p_x,L^*)
\end{equation}\\
where $p_x$ is a point on the ground truth, whose arc length to the first point is $x$ mm, $L^*$ is the normal section profile reconstructed by our method, $\mathop{d}(,)$ is the distance from a point to a curve.

\begin{table}[htbp]
	\centering
	\caption{The accuracy $Acc$ of the reconstructed normal section profile compared with the ground truth obtained the 3D scanner ATOS Core 185}
	\resizebox{\columnwidth}{!}{
	  \begin{tabular}{cccccc}
	  \toprule
	  No.   & Acc / mm & No.   & Acc / mm & No.   & Acc / mm \\
	  \midrule
	  1     & 0.067  & 11    & 0.064  & 21    & 0.081  \\
	  2     & 0.064  & 12    & 0.063  & 22    & 0.062  \\
	  3     & 0.069  & 13    & 0.056  & 23    & 0.064  \\
	  4     & 0.065  & 14    & 0.075  & 24    & 0.073  \\
	  5     & 0.077  & 15    & 0.068  & 25    & 0.059  \\
	  6     & 0.074  & 16    & 0.064  & 26    & 0.074  \\
	  7     & 0.076  & 17    & 0.061  & 27    & 0.060  \\
	  8     & 0.066  & 18    & 0.062  & 28    & 0.070  \\
	  9     & 0.075  & 19    & 0.065  & 29    & 0.060  \\
	  10    & 0.081  & 20    & 0.070  & 30    & 0.057  \\
	  \midrule
	  \multicolumn{6}{c}{Mean:   0.068 mm          Std: 0.007 mm          Min: 0.056 mm        Max: 0.081 mm} \\
	  \bottomrule
	  \end{tabular}%
	}
	\label{TABLE V}%
  \end{table}%

\mbox{Fig. \ref{Fig. 14}(c)} shows the point-wise errors along the reconstructed normal section profile. The reconstruction procedure was repeated 30 times and generates 30 curves, i.e. light color curves in the background in Fig. 11(c). The mean error of the 30 curves is shown as the red curve. We also visualize the mean point-wise errors along the profile via the heat map in \mbox{Fig. \ref{Fig. 14}(d)}, where the whole profile is divided into four segments. The mean error of the four segments range from 0.015 mm to 0.118 mm. as shown in Fig. 11(c).

For each time, we calculate an overall accuracy as \mbox{Eq. (\ref{formula (13)})}.
\begin{equation}\label{formula (13)}
	Acc = \big(\frac{1}{N_L} \sum_{\mathbf{p}\in L^*}\mathop{d}(\mathbf{p},L_0)\big)^{\frac{1}{2}}
\end{equation}
where $L^*$ is the normal section profile reconstructed by our method, $N_L$ is the number of points on $L^*$, $\mathbf{p}$ is a point on $L^*$, $L_0$ is the ground truth, $\mathop{d}(,)$ is the distance from a point to a curve.

The overall precision of the 30 repetitions is listed in \mbox{Table \ref{TABLE V}}. It shows that our algorithm can reach a mean precision of 0.068 mm and a good repeatability with the Std of 0.007 mm.

\section{Conclusion}\label{Sec_6}

This paper proposes a 3D reconstruction framework via multi-line structured light vision
for the normal section profile of 3D revolving geometrical structures.
In the framework, we propose a model to estimate the normal section profile and an iterative algorithm for optimization, which
allow the normal section profile to be reconstructed without constraints on the pose of sensor.
We conducted thorough real experiments on a partial 3D wheel of a train. The results demonstrate that
our framework with the proposed model and algorithm enable reliable and precise normal section profile
reconstruction of the 3D revolving structures. It achieves a mean precision of $0.068mm$ and
excellent repeatability with a standard deviation of $0.007mm$. The algorithm is robust to varying pose and position variations of the sensor, which greatly increases the flexibility of normal section profile reconstruction via a multi-line structured light sensor.

Note that the proposed model and algorithm can be generalized to any similar 3D revolving geometrical primitives,
and thus this 3D reconstruction scheme has good applicability in the application cases of industrial or manufacturing communities, e.g. measuring key geometrical parameters of a 3D revolving component through its normal section profile.
In the future, we could extend the proposed algorithm by integrating temporal information to 3D dynamic reconstruction of geometrical components in online application scenarios.

\section{Acknowledgment}\label{Sec_6}

The work is supported by The National Key Research and Development Program of China under Grant No.
2018YFB2003802 and National Nature Science Foundation of China (NSFC) under Grant No. 61906004.


\balance
\bibliographystyle{Bibliography/IEEEtranTIE}
\bibliography{Bibliography/BIB_xx-TIE-xxxx} 

\begin{thebibliography}{10}
\providecommand{\url}[1]{#1}
\csname url@samestyle\endcsname
\providecommand{\newblock}{\relax}
\providecommand{\bibinfo}[2]{#2}
\providecommand{\BIBentrySTDinterwordspacing}{\spaceskip=0pt\relax}
\providecommand{\BIBentryALTinterwordstretchfactor}{4}
\providecommand{\BIBentryALTinterwordspacing}{\spaceskip=\fontdimen2\font plus
\BIBentryALTinterwordstretchfactor\fontdimen3\font minus
  \fontdimen4\font\relax}
\providecommand{\BIBforeignlanguage}[2]{{%
\expandafter\ifx\csname l@#1\endcsname\relax
\typeout{** WARNING: IEEEtran.bst: No hyphenation pattern has been}%
\typeout{** loaded for the language `#1'. Using the pattern for}%
\typeout{** the default language instead.}%
\else
\language=\csname l@#1\endcsname
\fi
#2}}
\providecommand{\BIBdecl}{\relax}
\BIBdecl

\bibitem{gronskov1994apparatus}
L.~Gronskov, ``Apparatus for the scanning of a profile and use hereof,'' Oct.~4
  1994, uS Patent 5,351,411.

\bibitem{zhang2018high}
S.~Zhang, ``High-speed 3d shape measurement with structured light methods: A
  review,'' \emph{Optics and Lasers in Engineering}, vol. 106, pp. 119--131,
  2018.

\bibitem{van2016real}
S.~Van~der Jeught and J.~J. Dirckx, ``Real-time structured light profilometry:
  a review,'' \emph{Optics and Lasers in Engineering}, vol.~87, pp. 18--31,
  2016.

\bibitem{abu2016simple}
B.~A. Abu-Nabah, A.~O. ElSoussi, and A.~E.~K. Al~Alami, ``Simple laser vision
  sensor calibration for surface profiling applications,'' \emph{Optics and
  Lasers in Engineering}, vol.~84, pp. 51--61, 2016.

\bibitem{dong2019situ}
Y.~Dong and B.~Pan, ``In-situ 3d shape and recession measurements of ablative
  materials in an arc-heated wind tunnel by uv stereo-digital image
  correlation,'' \emph{Optics and Lasers in Engineering}, vol. 116, pp. 75--81,
  2019.

\bibitem{molleda2016profile}
J.~Molleda, R.~Usamentiaga, {\'A}.~F. Millara, D.~F. Garc{\'\i}a, P.~Manso,
  C.~M. Su{\'a}rez, and I.~Garc{\'\i}a, ``A profile measurement system for rail
  quality assessment during manufacturing,'' \emph{IEEE Transactions on
  Industry Applications}, vol.~52, no.~3, pp. 2684--2692, 2016.

\bibitem{wang2018distortion}
C.~Wang, Y.~Li, Z.~Ma, J.~Zeng, T.~Jin, and H.~Liu, ``Distortion rectifying for
  dynamically measuring rail profile based on self-calibration of multiline
  structured light,'' \emph{IEEE Transactions on Instrumentation and
  Measurement}, vol.~67, no.~3, pp. 678--689, 2018.

\bibitem{torabi2018high}
M.~Torabi, S.~M. Mousavi, and D.~Younesian, ``A high accuracy imaging and
  measurement system for wheel diameter inspection of railroad vehicles,''
  \emph{IEEE Transactions on Industrial Electronics}, vol.~65, no.~10, pp.
  8239--8249, 2018.

\bibitem{pan2019site}
X.~Pan, Z.~Liu, and G.~Zhang, ``On-site reliable wheel size measurement based
  on multisensor data fusion,'' \emph{IEEE Transactions on Instrumentation and
  Measurement}, 2019.

\bibitem{zhang2017detection}
J.~Zhang, Z.~Yang, Y.~Zhang, and Z.~Xing, ``Detection of wheel tread wear based
  on laser displacement sensor,'' in \emph{International Conference on
  Electrical and Information Technologies for Rail Transportation}, pp.
  399--408.\hskip 1em plus 0.5em minus 0.4em\relax Springer, 2017.

\bibitem{nayebi2010method}
K.~Nayebi, ``Method, apparatus, and system for non-contact manual measurement
  of a wheel profile,'' May~11 2010, uS Patent 7,715,026.

\bibitem{mian2009wheel}
Z.~F. Mian, J.~C. Mullaney, R.~MacAllister, and W.~Peabody, ``Wheel measurement
  systems and methods,'' Jan.~20 2009, uS Patent 7,478,570.

\bibitem{medianu2014system}
S.~O. Medianu, G.~A. Rimbu, D.~Lipcinski, I.~Popovici, and D.~Strambeanu,
  ``System for diagnosis of rolling profiles of the railway vehicles,''
  \emph{Mechanical Systems and Signal Processing}, vol.~48, no. 1-2, pp.
  153--161, 2014.

\bibitem{wu2010novel}
B.~Wu, T.~Xue, T.~Zhang, and S.~Ye, ``A novel method for round steel
  measurement with a multi-line structured light vision sensor,''
  \emph{Measurement Science and Technology}, vol.~21, no.~2, p. 025204, 2010.

\bibitem{xing2016online}
Z.~Xing, Y.~Chen, X.~Wang, Y.~Qin, and S.~Chen, ``Online detection system for
  wheel-set size of rail vehicle based on 2d laser displacement sensors,''
  \emph{Optik-International Journal for Light and Electron Optics}, vol. 127,
  no.~4, pp. 1695--1702, 2016.

\bibitem{cheng2016novel}
X.~Cheng, Y.~Chen, Z.~Xing, Y.~Li, and Y.~Qin, ``A novel online detection
  system for wheelset size in railway transportation,'' \emph{Journal of
  Sensors}, vol. 2016, 2016.

\bibitem{sun2009universal}
J.~Sun, G.~Zhang, Q.~Liu, and Z.~Yang, ``Universal method for calibrating
  structured-light vision sensor on the spot,'' \emph{J. Mech. Eng.}, vol.~45,
  no.~03, pp. 174--177, 2009.

\bibitem{besl1992method}
P.~J. Besl and N.~D. McKay, ``Method for registration of 3-d shapes,'' in
  \emph{Sensor fusion IV: control paradigms and data structures}, vol. 1611,
  pp. 586--606.\hskip 1em plus 0.5em minus 0.4em\relax International Society
  for Optics and Photonics, 1992.

\bibitem{chen1992object}
Y.~Chen and G.~Medioni, ``Object modelling by registration of multiple range
  images,'' \emph{Image and vision computing}, vol.~10, no.~3, pp. 145--155,
  1992.

\bibitem{donoso2017icp}
F.~Donoso, K.~J. Austin, and P.~R. McAree, ``How do icp variants perform when
  used for scan matching terrain point clouds?'' \emph{Robotics and Autonomous
  Systems}, vol.~87, pp. 147--161, 2017.

\bibitem{pottmann2009integral}
H.~Pottmann, J.~Wallner, Q.-X. Huang, and Y.-L. Yang, ``Integral invariants for
  robust geometry processing,'' \emph{Computer Aided Geometric Design},
  vol.~26, no.~1, pp. 37--60, 2009.

\bibitem{cheng1995mean}
Y.~Cheng, ``Mean shift, mode seeking, and clustering,'' \emph{IEEE transactions
  on pattern analysis and machine intelligence}, vol.~17, no.~8, pp. 790--799,
  1995.

\bibitem{zhang2000flexible}
Z.~Zhang, ``A flexible new technique for camera calibration,'' \emph{IEEE
  Transactions on pattern analysis and machine intelligence}, vol.~22, 2000.

\bibitem{ATOSCore185}
GOM, ``the atos core 185 scanner,'' accessed 25 November 2019.
  https://www.gom.com/metrology-systems/atos/atos-core.html.

\end{thebibliography}


\end{document}